\documentclass[preprint2,twoside]{hwo}

\usepackage{orcidlink}
\usepackage[symbol]{footmisc}

\usepackage{siunitx}
\let\captionbox\relax
\usepackage{subcaption}
\usepackage{wrapfig}
\usepackage{enumitem}
\usepackage{mathtools}

\DeclarePairedDelimiter\abs{\lvert}{\rvert}

\DeclareCaptionLabelFormat{paren}{#2)}
\input{hwo.h}

\makeatletter
\let\orig@maketitle\maketitle
\renewcommand{\maketitle}{
  \begingroup
    \renewcommand{\thefootnote}{\fnsymbol{footnote}}
    \orig@maketitle
  \endgroup
  \setcounter{footnote}{0}
}
\makeatother

\begin{document}

\title{\textbf{\LARGE Optical Design Pathways to Fluidic Space-Assembled Reflectors \& \\ Dual-Configuration Spectrographs for Characterizing Exo-Earths}}
\author {\textbf{\large Enrico Biancalani\protect\footnote{Virtual correspondence address: \href{mailto:ebiancalani94@gmail.com}{ebiancalani94@gmail.com}.} \! \orcidlink{0000-0002-6137-0342}$^{1, 2}$ \(\vert \vert\) Edward Balaban$^3$\(\vert\)\ref{FLUTE}., Ruslan Belikov$^3$\(\vert\)\ref{FLUTE}., Eduardo Bendek$^3$\(\vert\)\ref{FLUTE}., Valeri Frumkin$^4$\(\vert\)\ref{FLUTE}., Israel Gabay$^5$\(\vert\)\ref{FLUTE}., Guangjun Gao$^2$\(\vert\)\ref{ExoSpec}., Qian Gong$^2$\(\vert\)\ref{ExoSpec}.-\ref{FLUTE}., Christine Gregg$^3$\(\vert\)\ref{FLUTE}., Tyler Groff$^2$\(\vert\)\ref{ExoSpec}., Joseph Howard$^2$\(\vert\)\ref{FLUTE}., Omer Luria$^6$\(\vert\)\ref{FLUTE}., Michael McElwain$^2$\(\vert\)\ref{ExoSpec}.-\ref{FLUTE}., Lee Mundy$^1$\(\vert\)\ref{FLUTE}., Rachel Ticknor$^3$\(\vert\)\ref{FLUTE}., Sylvain Veilleux$^1$\(\vert\)\ref{ExoSpec}.-\ref{FLUTE}., \& Neil Zimmerman$^2$\(\vert\)\ref{ExoSpec}.-\ref{FLUTE}.}}
\affil{\small\it $^1$University of Maryland, College Park, USA. $^2$NASA Goddard Space Flight Center, USA. $^3$NASA Ames Research Center, USA. \\ $^4$Boston University, USA. $^5$Cornell University, USA. $^6$Technion – Israel Institute of Technology.}

\begin{abstract}
\textbf{\textit{Fluidic Telescopes}\(\vert\)\ref{FLUTE}.} We present a conceptual framework for optically designing space-assembled telescopes whose primary mirror is formed \textit{in situ} via the enabling, scale-invariant technology of fluidic shaping. In-space assembly of optical reflectors can solve light-gathering aperture scaling, which currently limits space-borne optical telescopes. Our compass reduces the top-level optical design trade to three types of avenues---a fluidic pathway, a legacy one building upon the James Webb Space Telescope, and hybrid solutions---with a focus on exo-Earths. A primarily fluidic pathway leads, in the first place, to a post-prime-focus architecture. We apply this configuration to propose the tentative optical design for a \(\sim\!\)1-m technology demonstrator and pathfinder for fluidic-telescope apertures scaling up to many tens of meters in diameter. 

\textbf{\textit{Dual-Configuration Spectrographs}\(\vert\)\ref{ExoSpec}.} The Habitable Worlds Observatory (HWO) will be the first mission equipped for the high-contrast direct imaging and remote spectral characterization, in reflected starlight, of exo-Earths in our galactic neighborhood. We present a novel concept for a compact, dual-configuration HWO spectrograph tailored for a broad wavelength range covering at least 600\,--1000 nm. Our design can interchange dispersive elements via a slider mechanism while preserving the rest of the optical path, enabling both a spectral resolving power \(R\!\sim\)140 integral-field spectrograph and a single- or multi-object spectrograph with \(R\) on the order of 10\(^3\). Although \(R\!\sim\)140 is near-optimal for the \(O_2\) absorption \(A\)-band around 760 nm, higher values of \(R\) can be utilized with spectral cross-correlation matched-filter techniques to enhance, e.g., HWO's atmospheric characterization capabilities.
  \\
  \\
\end{abstract}

\vspace{2cm}

\setcounter{footnote}{0}
\renewcommand{\thefootnote}{\arabic{footnote}}

\section{Introduction}

Life as we know it---or can imagine it---is able to spring out in seemingly inhospitable environments, such as the bottom of an ocean around hydrothermal vents; perhaps on moons such as Enceladus in our Solar System \citep{Voyage2050}. 
However, so far, \textit{in situ} missions cannot reach other stars. Instead, the next generations of telescopes have a chance to remotely assess whether our galactic neighborhood harbors any habitable or inhabited worlds other than Earth. 
Arguably, the aiming point of a systematic search for alien life havens shall start from targets that reflect the habitat where humans have been developing: i.e., Earth-sized extrasolar worlds orbiting in the habitable zone of Sun-like stars, alias exo-Earths, which can possibly sustain surface liquid water. 

This quest calls for an unprecedented suite of novel technologies. In fact, Earth turns out to be on the order of \(10^{10}\) fainter than the Sun in visible reflection. Disentangling such a brightness contrast for an Earth twin can only be attained with an ultra-stable adaptive coronagraphic system---at the picometer level per wavefront control cycle \citep{JWST_Lessons}---from outer space, not hindered by Earth's atmospheric turbulence. In parallel, an \textit{ad hoc} space telescope should also be large enough to collect as many photons per minute as needed to disambiguate potential biosignatures from any background astrophysical and instrumental noise sources, for as distant exoplanetary systems as possible; and to angularly resolve the host star from its planetary companion. Moreover, the photon-starved high-contrast direct-imaging observation of exo-Earths may benefit from different modes of resolving the spectral fingerprints characteristic of the target's atmosphere or surface, trading some detection sensitivity for spectral resolution. A trade that, in turn, ties to the signal-to-noise capabilities of the detector, which is fed by the spectrograph, after starlight suppression. 

In this paper, Section \ref{FLUTE} explores the optical architecture trade space of space-assembled reflectors whose primary mirror is formed via the scale-invariant fluidic shaping technology. Then, Section \ref{ExoSpec} proposes a novel concept of multi-configuration, imaging spectrograph to get low- and moderate-resolution spectra for exo-Earth characterization.

\clearpage

\section{Fluidic Space Telescopes: Evolution vs Revolution}
\label{FLUTE}

\begin{figure}[ht]
\centering
\includegraphics[width=\columnwidth]{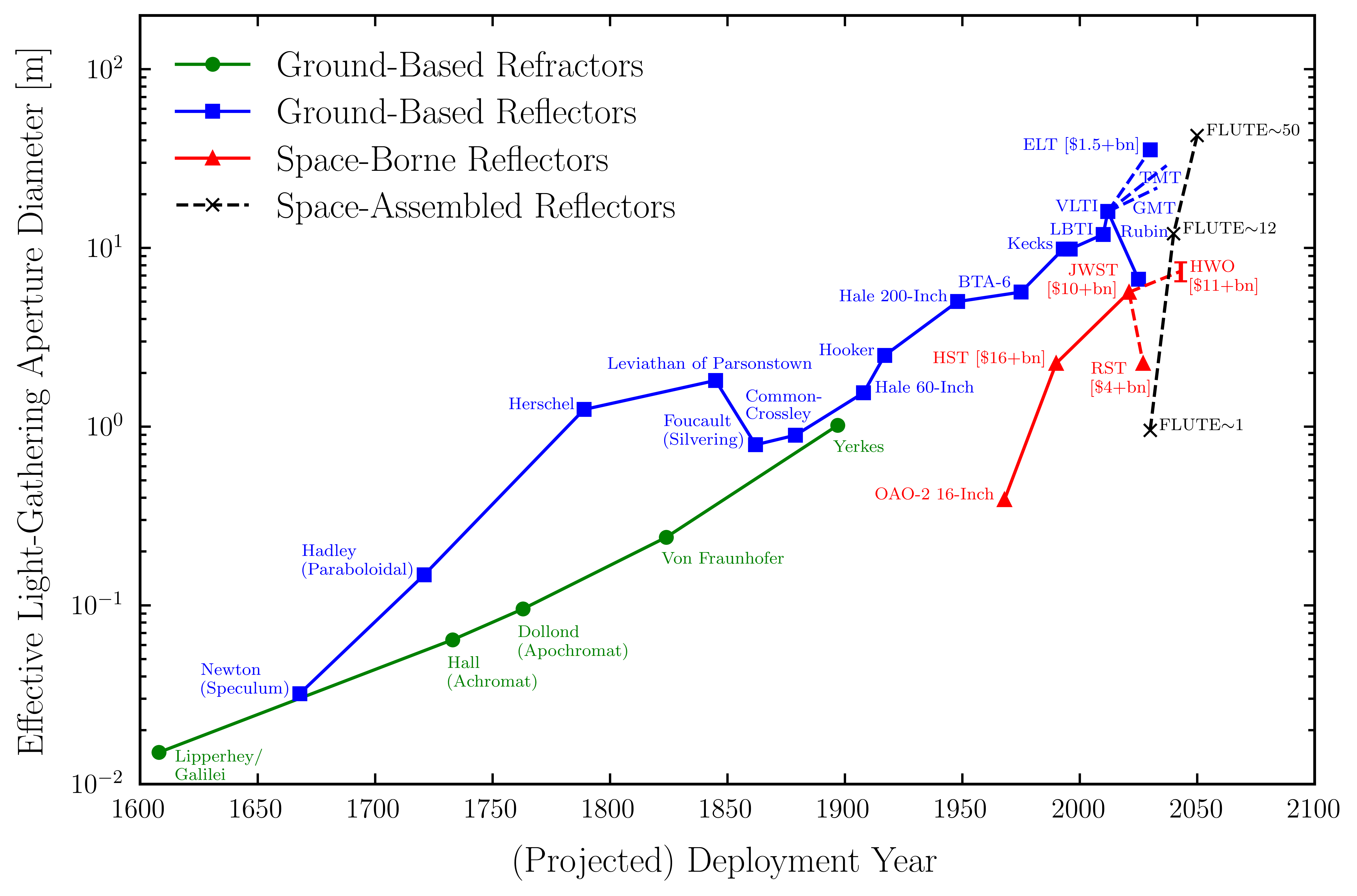}
\caption{\small Diachronic macro-evolution graph\protect\footnotemark{} of optical telescopes' effective primary aperture's diameter, i.e., accounting for secondary-optics geometrical obscurations. Dashed lines indicate potential or future telescopes. Annotated acronyms and the like for modern telescopes: OAO-2 = Orbiting Astronomical Observatory 2; BTA = Bolshoi Teleskop Alt-azimutalnyi (romanized Cyrillic script, meaning ``Large Alt-azimuth Telescope'' in Russian); HST = Hubble Space Telescope; LBTI = Large Binocular Telescope Interferometer; VLTI = Very Large Telescope Interferometer; JWST = James Webb Space Telescope; RST = Roman Space Telescope; ELT = Extremely Large Telescope; FLUTE = Fluidic Telescope (with \(\sim\!\!1\)-m, \(\sim\!\!12\)-m, and \(\sim\!\!50\)-m primary apertures); GMT = Giant Magellan Telescope; TMT = Thirty Meter Telescope; HWO = Habitable Worlds Observatory. The minimum cost estimates, adjusted for inflation to recent years, are sourced from the internet.
}
\label{Macro-Evolution of Optical Telescopes}
\end{figure}

Let us consider the macro-evolution of optical telescopes since their conception, as represented within Fig. \ref{Macro-Evolution of Optical Telescopes}. This selection indicates an overall diachronic trend: their effective light-gathering aperture's diameter---i.e., accounting for obscurations---is scaling roughly exponentially since the times of Hans Lipperhey \citep{78d0c3c3f84b4e509b11256f872d626f} and Galileo Galilei for ground-based telescopes \citep{2004PASP..116...77R}. Another key insight comes from the distinction between space-borne and space-assembled optical reflectors and their (projected) trajectories. The former are a direct evolution of ground-based telescopes that are launched into outer space: e.g., the James Webb Space Telescope, commonly known as JWST, which had to be folded up to fit into its rocket's payload nose-cone fairing. The latter potential branch is growing out of the Fluidic Telescope---alias FLUTE---project \citep{FLUTE_Project}, which aims at in-space assembling (at least) the primary mirror via the novel approach of ``fluidic shaping'', recently introduced by \citet{Frumkin_Bercovici_2021}. 

The fork in the development of space telescopes responds to a major technological challenge: i.e., the lack of a technology capable of bringing large-class primary mirrors to outer space that keep up with the overall scaling trend. In fact, the space-borne branch in Fig. \ref{Macro-Evolution of Optical Telescopes} appears to be plateauing, not considering the Nancy Grace Roman Space Telescope (abbreviated RST). The reason for this exclusion is that RST incorporates a primary mirror---twin of the Hubble Space Telescope (HST) one---donated with other core optics to NASA by the US National Reconnaissance Office \citep{2012Natur.490...16H}. In addition to overcoming key development obstacles, such as multi-decadal gestation periods and high marching-army costs, the goal is to leverage outer space at best while rivaling the capabilities of the upcoming generation of ground-based extremely large telescopes, such as the eponymous ELT. In turn, reflectors seem to be reaching the single-aperture limit of what is practicable on Earth under the effect of gravity, akin to the fate of the ground-based refractors after the Yerkes Observatory's pinnacle. 

Besides, the ``seeing'' effect resulting from Earth's atmospheric turbulence hinders the astrophysical discoveries attainable from the ground. E.g., coronagraphs atop the ELTs will be limited to a planet-to-star luminous flux ratio on the order of \(10^{-7}\) for the habitable zone of nearby \(M\)-dwarf stars, even with the aid of extreme adaptive optics and post-processing recovery techniques \citep{2025ARA&A..63..179K}. This value is still far from the \(10^{-10}\) level required for exo-Earths around solar-type stars, to which HWO is going to be dedicated. The sky background flux is another external factor affecting the quality of ground-based observations: this comprises airglow and thermal emission, telluric extinction due to absorption and scattering, as well as light pollution from the growing satellite ``mega-constellations''.

\newpage

\footnotetext{The latest version of this telescope scaling graph resides \href{https://github.com/WhiteDogos/Macro-Evolution-Telescopes}{\textbf{here}}.}

\subsection{Pinning, No Spinning: Optics by Fluidic Shaping}

Let us now consider the formula for the capillary length of a fluid-fluid interface \citep{Frumkin_Bercovici_2021}:

\vspace{-5pt}

\begin{equation}
l_c = \sqrt{\frac{\gamma}{|\Delta \rho| \, g}},
\label{Capillary Length Formula}
\end{equation}

\vspace{-1pt}

\noindent
where \(\gamma\) denotes the surface tension at the interface, \(\Delta \rho\) stands for the average mass-density difference between the fluids (or liquids, more specifically, here), and \(g\) is the standard gravitational acceleration. When a bead of dew on a leaf has a diameter much smaller than \(l_c\), the droplet---naturally drawn to its minimum energy state---tends to remain spherical, with molecular surface roughness. Similarly, a blob of liquid injected into outer space will eventually relax to a spherical shape in the ideal absence of external net forces. 
In general, given a finite \(\gamma\) in Eq. \ref{Capillary Length Formula}, \(l_c\) goes to infinity---the main condition for scale invariance---if either \(\Delta \rho\) goes to 0, which brings the system into a neutral buoyancy condition; or if \(g\) goes to 0, which leads to a micro-gravity condition. This means that we can possibly leverage outer space to assemble larger optical components than possible on Earth, where they would sag under gravity. 

\clearpage

The history of liquid mirrors for telescopic applications can be loosely traced back to Isaac Newton's rotating bucket experiment, although their potential has remained latent until few decades ago, mainly due to practical hurdles related to the spinning liquid \citep{2025AcAau.230...30C}. 
Essentially, the novelty of fluidic shaping consists in pinning a determined amount of liquid with some optical properties to a bounding frame. In the appropriate environment, surface tension governs the process of optical assembly, providing a naturally smooth surface within that boundary. 
\begin{wrapfigure}[12]{l}[0cm]{0.47\columnwidth}
    \vspace{-.4cm}
    \begin{minipage}{\linewidth}
        \centering
        \includegraphics[width=\columnwidth]{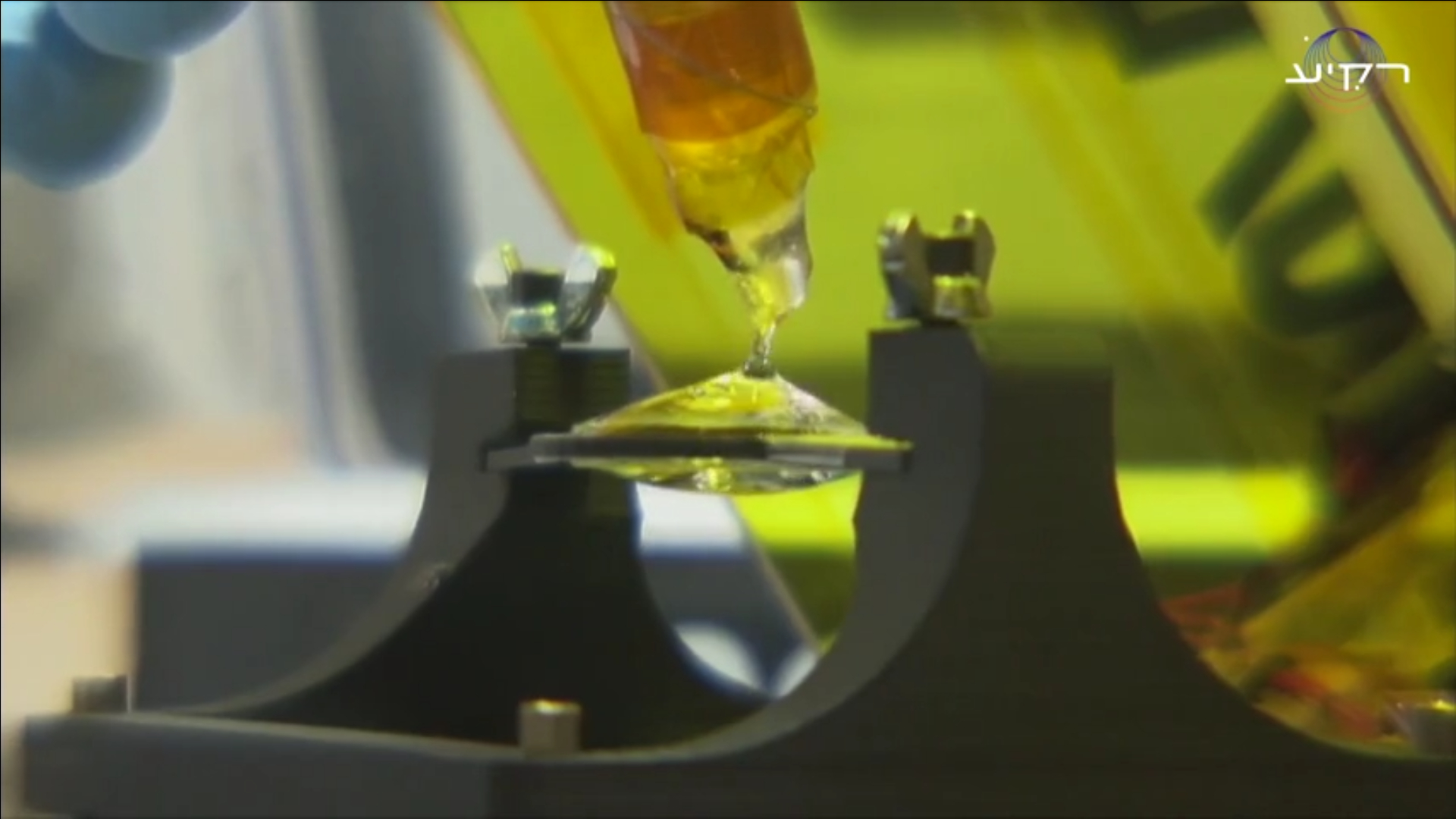}
        \caption{\small Frame of the video that displays the first lens that was fabricated in outer space, aboard the ISS, thanks to fluidic shaping \citep{Bercovici_Fluidic_Technologies_Lab}.}
        \label{ISS Fluidic Lens Experiment}
    \end{minipage}
    \vspace{-.5cm}
\end{wrapfigure}
If an amount of transparent liquid gets injected that exceeds the level of a circular boundary strip, the product is a biconvex spherical lens (symmetrical on both sides of the boundary). In April 2022, the first lens was created in outer space, aboard the International Space Station (alias ISS), via fluidic shaping, as shown in Fig. \ref{ISS Fluidic Lens Experiment}. 
Removing liquid from the system below the boundary's level results in a biconcave spherical lens. Ideally, if a hypothetical transparent planar surface is inserted through the middle of the strip, the lens turns into two plano-concave spherical lenses facing each other (via the planar surface). At this point, adding reflectivity to the two concave sides of the optical system---e.g., by infusing reflective nanoparticles---creates two concave mirrors adjacent to each other back-to-back. The roughness of the mirrors' surface is independent of the roughness characterizing the planar surface beneath, provided that the liquid layer is thick enough with respect to that. Actually, the shape of the liquid mirrors does not change if the planar surface is replaced, e.g., by a spherical cap, so long as the boundary conditions are preserved. Therefore, it will be convenient to accompany the curvature of the liquid mirrors with a floor---its material and structure do not matter for the sake of this mental experiment---a few millimeters beneath their surfaces, to save on the amount of liquid employed and to minimize the dynamical perturbations in the liquid itself. 
Overall, both refractive and reflective optical components can be constructed by pinning a liquid polymer to a geometrical boundary and controlling the volume of injected liquid; which can be solidified via ultraviolet curing, if necessary. 

For context, it is important to note that news about another kind of liquid telescope first made its way through the ears of the knowledgeable astronomer. A relatively cheap zenith telescope with a primary mirror generated by continuously and precisely spinning a liquid into a paraboloid of revolution. E.g., the cost of the 6-m Large Zenith Telescope (LZT), which was completed in 2003, operational since 2005, and decommissioned in 2016, amounted to less than one million US dollars \citep{2007PASP..119..444H}, roughly one fiftieth of a similar-sized telescope \citep{2004SPIE.5489..563V}---the drawback being that it could only observe at zenith. It is relevant here to notice that the Defense Advanced Research Projects Agency (DARPA) has recently established a program called ``Zenith'' \citep{2024SPIE13100E..37N} to overcome the limitations of a moving liquid primary in creating and operating liquid-mirror telescopes \citep{2024SPIE13131E..09B}. 

Fluidic shaping of telescopes' primary mirror can enable the technological paradigm shift to space-assembled observatory architectures, which can be strategically scaled up at a lower cost in an iterative fashion \citep{2023SPIE12677E..0DD}. In fact, fluidic space-assembled telescopes could leverage at best the capabilities offered by SpaceX's Starship in terms of payload lift, volume, and launch cadence. Whether extraterrestrial biosignatures \citep{2018AsBio..18..663S} are found over the next decades or not, the ongoing exoplanet discovery revolution is unfolding (almost) exponentially, with an empirical doubling time of few years. To date, we have just passed the total number of 6,000 confirmed exoplanets\footnotemark{}. A fact that should drive long-term plans of multiple observatories to search deeper and systematically from outer space---including the Moon. So, why not speculate about telescope constellations at the Sun-Earth Lagrangian point 2 (SEL2), which does not have the disadvantage of orbital visibility due to cyclical Earth occultations, much in the same way as there are satellite constellations nowadays; perhaps linking the telescopes for aperture-synthesis imaging \citep{2010AsBio..10..121S} like the Large Interferometer for Exoplanets concept, known as LIFE \citep{2022A&A...664A..21Q}. 

The (disclosed) state of the art boasts the lens fabrication experiment aboard the ISS shown in Fig. \ref{ISS Fluidic Lens Experiment}. Apart from this, lens manufacturing in a neutral buoyancy environment---by matching the density of the optical liquid with that of an immersion liquid, on Earth---was proven to reproduce the gamut of possible free-form fluidic lenses \citep{2021Optic...8.1501E, 2024arXiv240604937C, 10.1145/3680528.3687584}; slight deviations from neutral buoyancy being required for asphericity. A fact that, incidentally, could solve lens manufacturing for eyeglasses in low-resource settings \citep{bf8746f74855402c9e89b39970cca0f8} on Earth and in view of long-duration crewed space missions (e.g., to Mars). A temporary reduced-gravity condition can be attained on parabolic flights, where both lenses \citep{2023npjMG...9...74L} and mirrors were created. The reflectivity of the mirror surface can be enhanced by utilizing a gallium alloy or an ionic liquid coated with reflective nano-particles \citep{c58b1e21c1ed4e4d8bea908204a7baf6}. 
The next logical milestone is the first assembly of a fluidic mirror in orbital free-fall around Earth. Future prospects involve a system technology demonstrator and pathfinder for a fluidic space-assembled telescope. 

\footnotetext{According to, e.g., the \href{https://exoplanetarchive.ipac.caltech.edu/}{NASA Exoplanet Archive}.}

\clearpage

\subsection{FLUTE\(\sim\)1 \& Beyond: An Architectural Compass}

Fluidic shaping of telescopes' primary mirror offers a potential scale-invariant solution to leverage outer space at best and to overcome the limits imposed by gravity on Earth, thus enabling space observatories to eventually close the scaling gap with their ground-based counterparts. Large light-gathering apertures bring in multiple benefits, which make the spread---besides the optical qualities---of a telescope's primary its most important characteristic; namely: 

\vspace{.1cm}

\begin{itemize}[leftmargin=*, nosep]
    \item Better angular resolution, leading to a smaller telescope direct-imaging inner working angle (IWA), inversely proportional to the extent of the light-gathering aperture. This also allows to probe companions closer to the host star or farther exoplanetary systems. For reference, an Earth twin 10 parsecs away from us would orbit at an apparent angular separation of \(\sim\!\!0.1\) arcseconds from its Sun-like host star. E.g., doubling the telescope diameter gives access to the same Earth twin at double the distance. 
    \item Much lower exposure time; or, conversely, much higher imaging speed. It approximately goes with the inverse fourth power of the diameter, for diffraction-limited point sources in background-limited regime. Leveraging such a scaling law is fundamental for capturing fast or diurnal exoplanet cycles in their atmosphere or on their surface. E.g., doubling the light-gathering aperture's diameter translates into a \(\sim\!\)16-fold time gain: a cumulative exposure of \(\sim\!\)16 days would only take a single day. 
    \item Higher light-gathering power, proportional to the telescope's effective light-gathering area, which enhances the photometric sensitivity---treating the losses within the telescope as constant with respect to the telescope's diameter. E.g., doubling the telescope's effective diameter increases the survey volume by a factor of 8. Alternatively, one could deploy integral-field spectrographs with higher spectral resolving power, \(R\!\vcentcolon=\!\frac{\lambda_{\text{ref.}}}{\abs{\Delta\lambda}}\!\equiv\!\frac{E_{\text{ref.}}}{\abs{\Delta E}}\), where \(\abs{\Delta\lambda}\) and \(\abs{\Delta E}\) denote the spectrograph's resolution in photon wavelength and energy, respectively. In fact, these need more signal for further spectral dispersion given the photon-starved exo-Earth characterization regime. One could do more efficient wavefront sensing and control system downstream, which is capable of compensating for the brighter quasi-static speckle field emerging from the mutual interference of coherent wavefronts of residual starlight leaked through the coronagraph and scattered throughout the telescope, in concert with post-processing techniques; from differential to machine-learning-driven differentiable-rendering techniques \citep{2018ARA&A..56..315G, 2025arXiv250101912F}. A higher number of collected photons also allows for the relaxation of coronagraphs' throughput curves and of tolerance requirements for a starshade: a revolutionary sunflower-shaped spacecraft origami-unfolded in outer space and flying in formation with the telescope, which would maximize the exo-Earth spectral characterization return \citep{2020arXiv200106683G} via broadband destructive interference. Given a certain starlight suppression level and IWA, there is a minimum design point in terms of starshade diameter and its distance from the telescope, both of which parameters scale linearly with the telescope's primary diameter \citep{2009SPIE.7440E..13G}. 
\end{itemize}

\noindent
In light of this, the design principles and best practices derived from the lessons learned by developing JWST offer some reference guidance on how to approach the optical design of FLUTE. In particular, the second design principle by \citet{JWST_Lessons}: ``Telescope evolution not revolution: scalable, verifiable telescopes building upon Webb (but with a baffle).'' In fact, a key question to ponder here is: to what extent do design principles like this apply to future flagship observatories in case a game-changing telescopic technology unexpectedly emerges and becomes viable? \\
To holistically design FLUTE, two pathways can be thought of as conceptually delimiting the optical architecture trade space: a revolutionary pathway based on the fluidic shaping technology of the primary mirror; and an established legacy pathway harnessing fluidic shaping to expand the capabilities of current observatories. Let us start by adopting two legacy design points building upon JWST in our design: 
\vspace{0.06cm}
\begin{itemize}[leftmargin=*, nosep]
    \item Aspherical primary mirror shape, appropriately modulating the topology of the deployable frame's supporting meshed floor \citep{10520999} to reduce the amount of liquid employed and minimize the liquid fluctuations. The added complexity with respect to the spherical cap is compensated for by the ease of reaching a diffraction-limited regime against static optical aberrations, mitigating the spherical aberration at the source and, thus, increasing the optical accuracy with lower systemic sensitivity. In outer space, active reshaping of the liquid sitting on the supporting floor is required to deviate from surfaces of constant mean curvature---the spherical cap, given a ring-shaped boundary---and create an aspherical fluidic mirror  (or free-form shape, in general). This could be achieved via electrostatic or magnetic actuation of the appropriate liquid deformable mirror \citep{2025AcAau.230...30C}; alternatively, by statically or dynamically exploiting the thermo-capillary effect (under investigation by Valeri Frumkin in our group). In addition, these methods could also drastically reduce the settling times and the amplitude of disturbances on the liquid mirror surface. 
    \item Fast-converging optical system, which reinforces the structural stability of the observatory. On the other hand, a slow optical configuration could mitigate static primary optical aberrations and allow for a more compact corrective optical element to avoid a smaller ray-optic caustics-enveloped region. In case of a large primary aperture, this would entail a free-flying corrective or instrument package---the added modularity favoring serviceability.
\end{itemize}

\clearpage

Table \ref{FLUTE Compass} summarizes the two aforementioned requirements within the space of top-level optical design trade-offs for FLUTE, classifying them along the evolutionary pathway \textbf{(a)}. This harnesses fluidic shaping to expand the capabilities of observatory concepts that see JWST as a direct legacy; and, thus, it is slated to more easily meet the favor of community stakeholders and be prioritized in the short and medium term. The revolutionary pathway \textbf{(b)} conceptually delimits the trade-off framework for telescopes built around fluidic shaping in order to maximize their scalability. More in detail, the optical trade space for FLUTE can be divided into a set of four choices going in categorical order: from the shape of the primary mirror (\textbf{1.}) and its surface continuity (\textbf{2.}); through the state of matter---liquid or solid---of the reflective material or its substratum, if any, during operation (\textbf{3.}); to the focal ratio between focal length of the primary mirror and outer effective diameter (\textbf{4.}), \(f/\# \vcentcolon = f/D\).

\vspace{-.23cm}

\begin{table}[ht]
\includegraphics[width=\columnwidth]{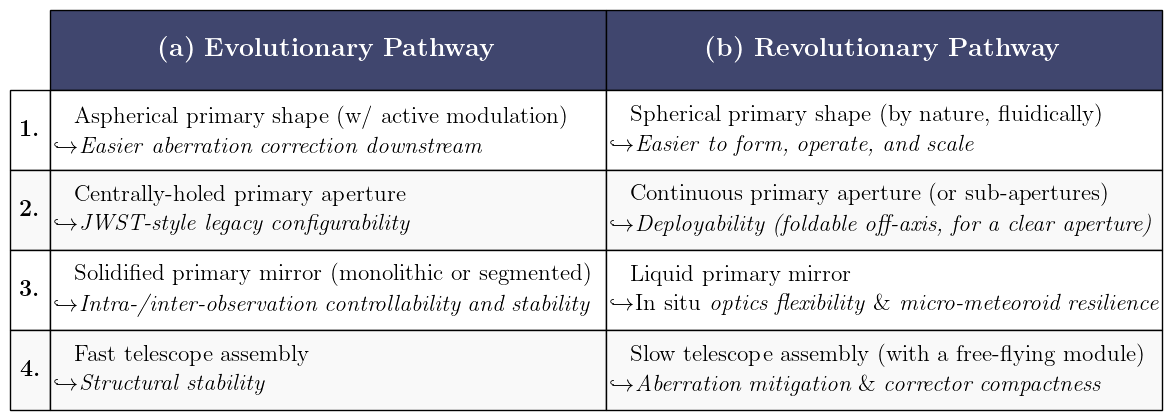}
\caption{\small Schematic conceptual framework for designing fluidic space-assembled reflectors. A design path can be variously combined along the four categories, between pathways \textbf{(a)} and \textbf{(b)}.
}
\label{FLUTE Compass}
\end{table}

\vspace{-.23cm}

\noindent
The current design points of choice along pathway \textbf{(b)} are: 
\begin{itemize}[leftmargin=*, nosep]
    \item Continuous surface of the primary aperture, to facilitate the optical liquid's initial deployment. This forces the rest of the optical train, as well as the instrumentation package, to be in front of the primary mirror or partially on the side of the telescope. E.g., an off-axis configuration was chosen for HWO's notional Exploratory Analytic Cases (EAC) 4 and 5\footnote{HWO's EAC-4 and EAC-5 were officially released at \href{https://www.stsci.edu/contents/events/stsci/2025/july/towards-the-habitable-worlds-observatory-visionary-science-and-transformational-technology}{HWO25}.} because a clear primary aperture is currently favorable for coronagraphs. An alternative solution is to collect light from the focal plane and transmit it to the back of the telescope along the support structure via optical waveguides. Conversely, a holed primary would open up the design space to more established optical architectures folded on-axis, such as JWST in Fig. \ref{JWST}. 
    \item Liquid primary mirror throughout operations---with thermal support. Crucially, this provides resilience to micro-meteoroids, reducing the need for a (mainly) protective barrel baffle. Furthermore, the liquid phase allows one to modify the telescope's optical power on the fly by changing the amount of liquid---which also helps to compensate for its evaporation. One may also imagine an adaptive fluidic corrective optical system---see, e.g., \citep{2025AcAau.230...30C}---coupled to the liquid primary mirror. 
\end{itemize}

\newpage

\begin{figure}[ht]
\centering

\subcaptionbox{FLUTE\(\sim\!\)1 architecture. \label{FLUTE-1 Optical Architecture}}[.42\columnwidth]{
  \raisebox{.45cm}{\includegraphics[height=.12\textheight]{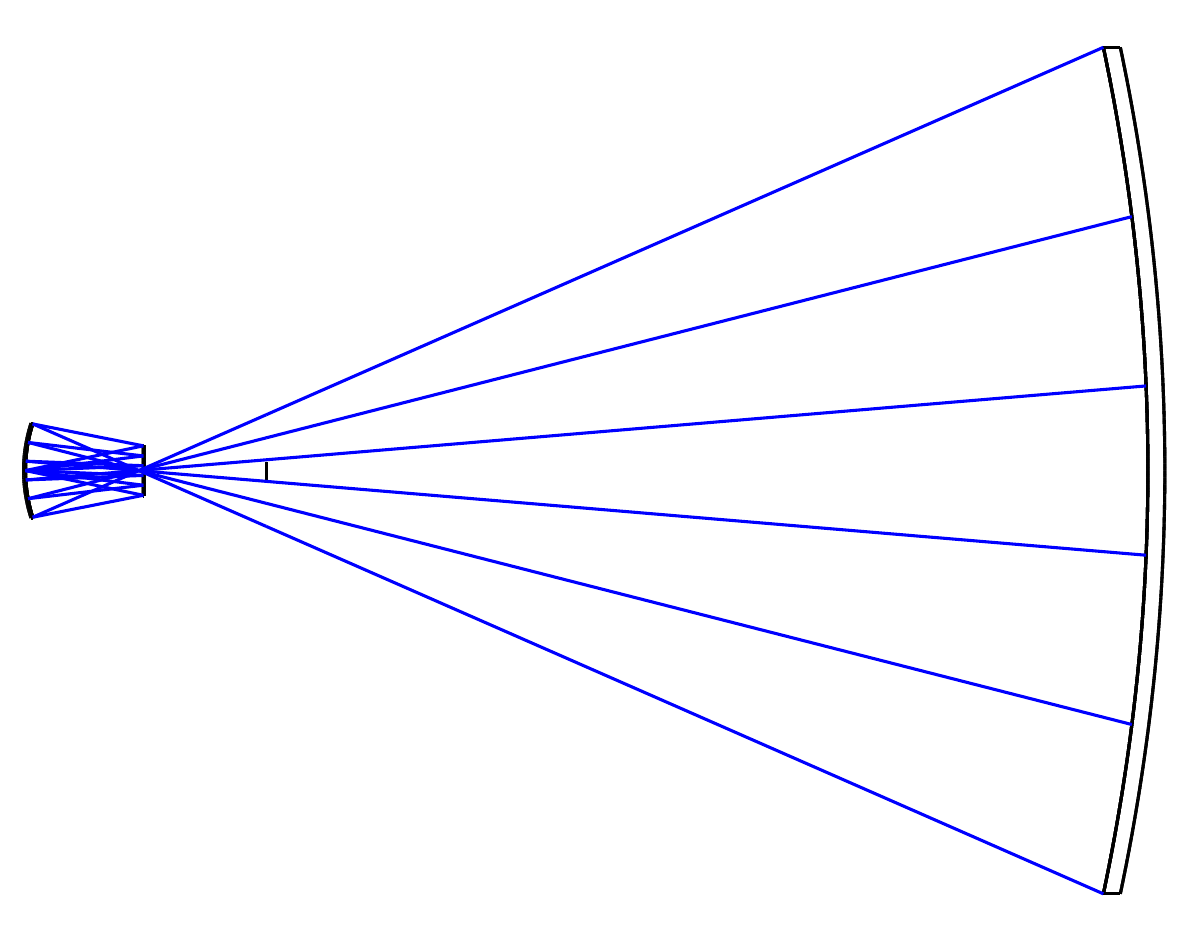}}}
\hfill
\subcaptionbox{FLUTE\(\sim\!\)1: opto-fluidic analysis. \label{FLUTE-1 Opto-Fluidic Analysis}}[.54\columnwidth]{
  \includegraphics[width=.54\columnwidth]{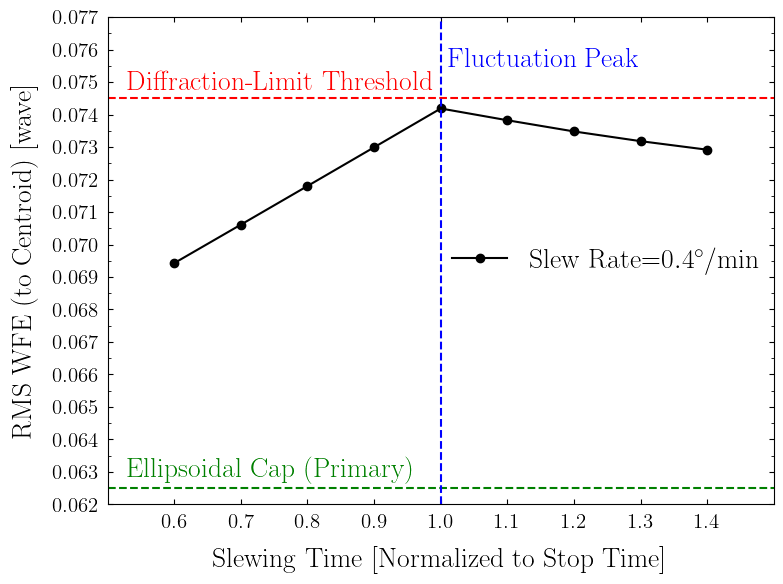}}
\vspace{-.45cm}
\caption{\small \ref{FLUTE-1 Optical Architecture}) Tentative post-prime-focus optical design layout---in Ansys Zemax OpticStudio---for FLUTE\(\sim\!\)1, with a prolate-ellipsoidal \(\sim\!\)1-m primary mirror on the right and, in order, central optical baffle and corrector on the left; rays trace the central field angle. The corrector consists of two holed conical mirrors; the instrumentation package would be connected on the back of the left-most mirror and pick up light from the internal focus. With an effective focal length of \(\sim\)2.4 m, it is diffraction-limited at 1 \SI{}{\micro\meter} over a circular field of view of \(\sim\)5 arcminutes in diameter, with \(\sim\)97\(\%\) fraction of rays unvignetted over this field of view. The diffraction limit here refers to a nominal root-mean-square (RMS) wavefront-error (WFE) threshold of \(\sim\!\!\lambda/13.4\) or \(\sim\)0.0745 waves, which is a standard WFE criterion (fully valid for an ideal clear aperture): i.e., 74.5 nm at a reference wavelength of 1 \SI{}{\micro\meter}. Fluidic shaping provides a sub-nanometric RMS surface roughness. \\ 
\hspace*{0.1cm} \ref{FLUTE-1 Opto-Fluidic Analysis}) Opto-fluidic analysis of FLUTE\(\sim\!\)1 undergoing a 10\(\si{\degree}\) slew maneuver, in terms of RMS WFE with respect to the optical centroid. At a slew rate of 0.4\(\si{\degree}\)/min, FLUTE\(\sim\!\)1 would be diffraction-limited throughout operations; faster slew rates may require active stabilization or solidification---for reference, JWST's requirement is \(\sim\)2\(\si{\degree}\)/min. With no fluctuations, the fiducial primary is an ellipsoidal cap. The main fluidic parameters defining the primary mirror's ionic liquid---according to Israel Gabay's purely fluidic underlying model---are: a 5-mm initial film thickness (which varies throughout operations), a surface tension of 50 mN/m, a mass density of 1700 kg/m\(^3\), and a kinematic viscosity of 10\(^{-4}\) m\(^2\)/s.}
\label{FLUTE-1}
\end{figure}

\begin{figure}[ht]
\centering

\subcaptionbox{JWST architecture. \label{JWST}}[0.47\columnwidth]{
  {\includegraphics[height=.12\textheight]{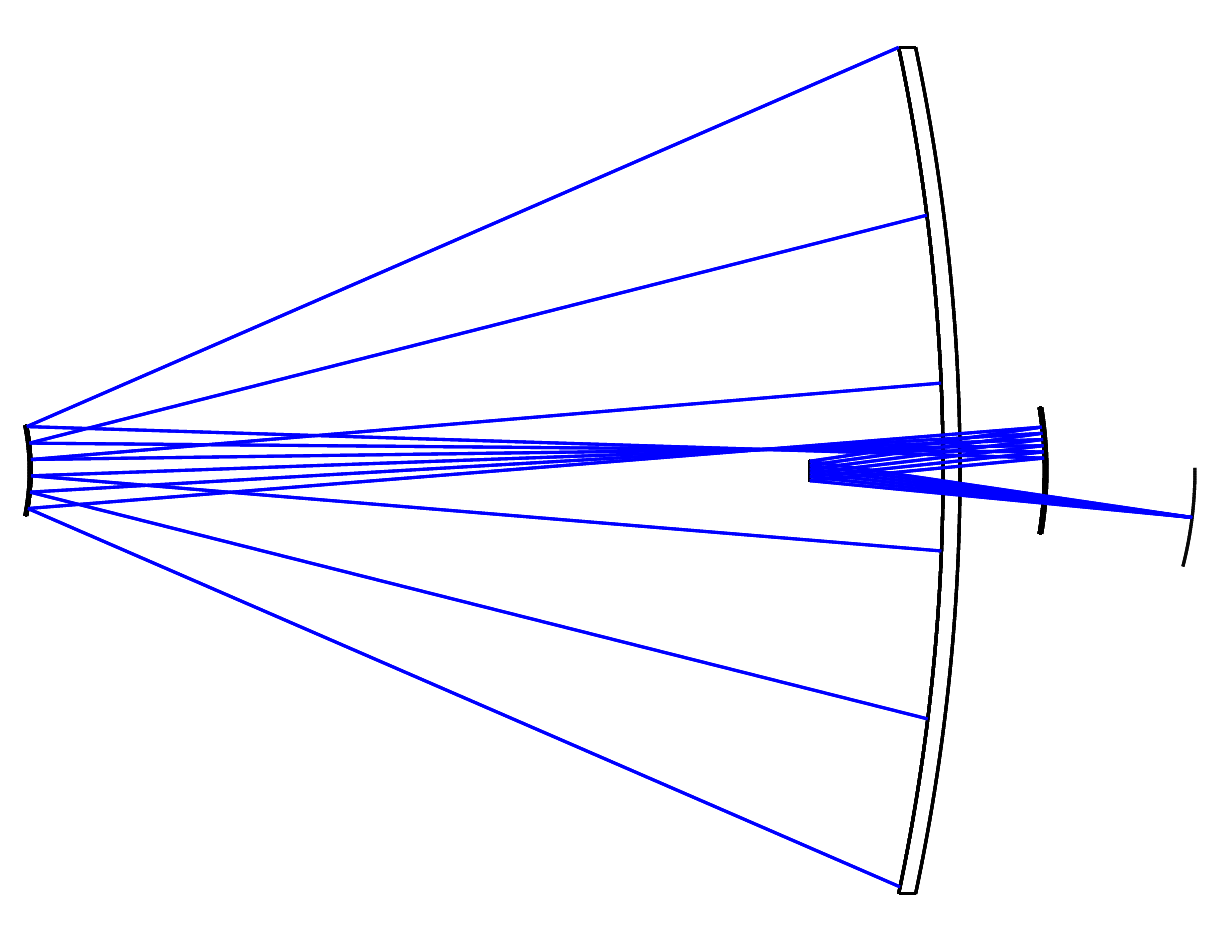}}}
\hfill
\subcaptionbox{FLUTE\(\sim\)50 architecture. \label{FLUTE-50}}[.47\columnwidth]{
  \includegraphics[height=.12\textheight]{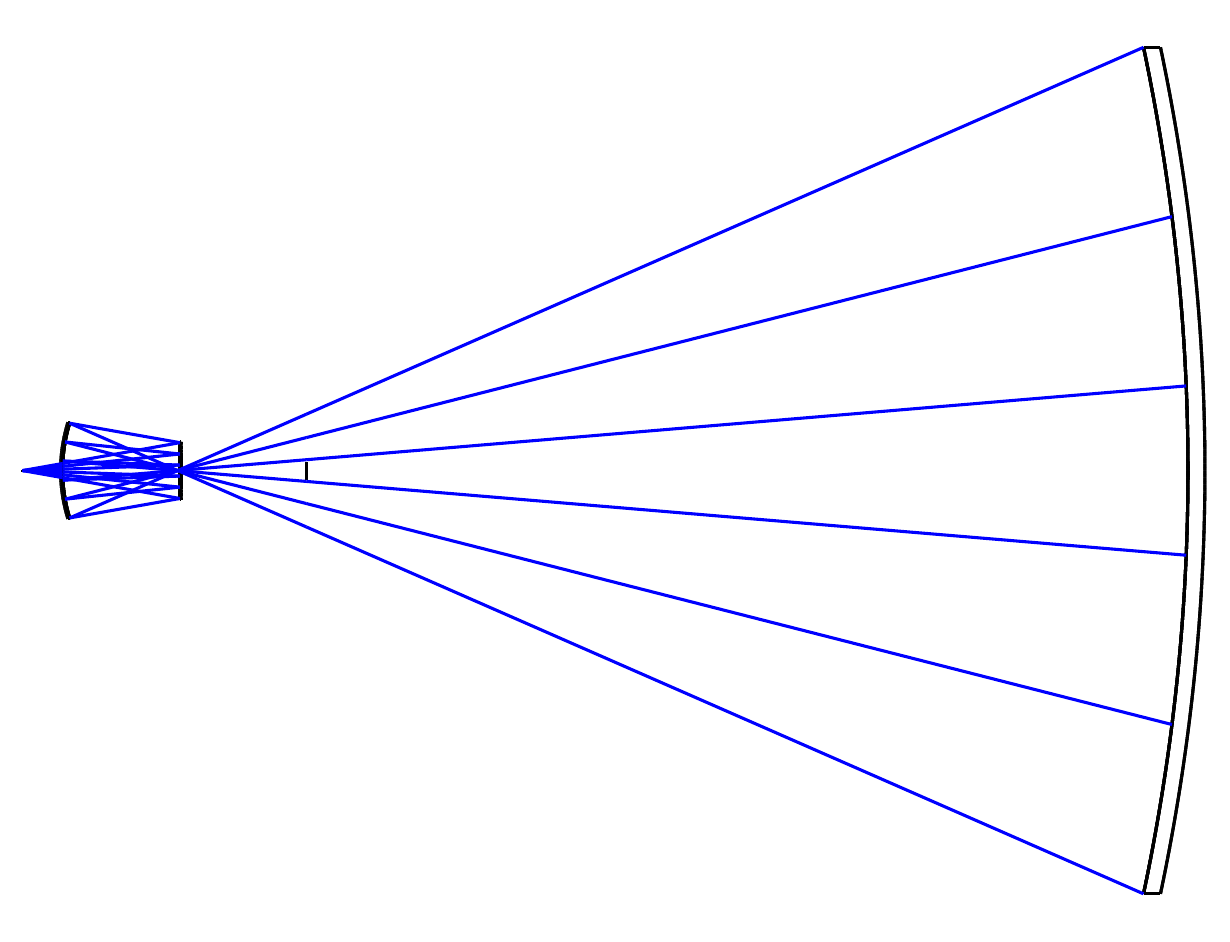}}

\caption{\small \ref{JWST}) Optical design layout of the \(\sim\)6.6-m-aperture JWST\protect\footnotemark{}, which has a folded three-mirror anastigmat configuration. As for the FLUTE designs in Fig. \ref{FLUTE-1 Optical Architecture} and in Fig. \ref{FLUTE-50}, JWST has a prolate-ellipsoidal primary mirror---shown without tessellation here---close to a paraboloidal shape, which is a perfect geometrical focuser on-axis at infinity: this deviation helps reduce the coma optical aberration for off-axis field angle points. JWST has an \(\sim\!\!f\)/20 optical system with a slight field angle bias (with respect to the central field, shown here) to accommodate the fine-steering mirror in front of the tertiary mirror. Its rectangular diffraction-limited field  of view at \(\sim\!\)1 \SI{}{\micro\meter} is \(\sim\!\)18\(\times\)9 arcminutes\(^2\). \\ 
\hspace*{0.1cm} \ref{FLUTE-50}) Tentative post-prime-focus optical design layout---in Ansys Zemax OpticStudio---for the \(\sim\)50-m-aperture FLUTE\(\sim\)50, with a prolate-ellipsoidal primary mirror, a central optical baffle, and the corrector; rays trace the central field angle. The corrector consists of two holed conical mirrors; the instrumentation package would be connected to the focus behind the secondary mirror. This is an \(\sim\!\!\!f\)/3 optical system with a much more compact optical folding than JWST in Fig. \ref{JWST}, which leaves the primary mirror continuous. On the other hand, it gives less margin of maneuver to correct aberrations for off-axis field angle points: the price to pay is on the diffraction-limited field of view (at 1 \SI{}{\micro\meter}), which is \(\sim\)0.8 arcminutes in diameter---a circular field of view---at the current iteration.}
\label{JWST vs FLUTE-50}
\end{figure}

By applying this hybrid combination of points from pathways \textbf{(a)} and \textbf{(b)}, Fig. \ref{FLUTE-1 Optical Architecture} proposes the tentative optical design for the prototypical version of FLUTE\(\sim\!\)1---this versioning refers to a primary mirror of \(\sim\!\)1 m in diameter. Fig. \ref{FLUTE-1 Opto-Fluidic Analysis} reports a preliminary opto-fluidic analysis of FLUTE\(\sim\!\)1, with a liquid primary, undergoing a slew maneuver. 

\vspace{-.3cm}

The strategic, programmatic goal is to mature the mission concept and the technology readiness level of its components in view of a next-generation NASA-led astrophysics flagship observatory at SEL2, with a notional \(\sim\)50-m primary aperture. Fig. \ref{JWST vs FLUTE-50} describes the tentative optical architecture of FLUTE\(\sim\)50 in comparison with JWST. Fluidic shaping is a scale-invariant approach that does not need folding of the primary optics if it is larger than the rocket fairing, unlike JWST. In turn, this reduces the system complexity; and makes alignment and adjustments simpler. It also reduces precautions and risk containment measures associated with any environmental rocket lift-off loads. 

\clearpage

\footnotetext{Reconstructed in Ansys Zemax OpticStudio thanks to \href{https://youtu.be/OE3ZowuvAJY}{\textbf{this webinar}}.}

\vspace{-.3cm}

However, a nimbler, intermediate-class \(\sim\!\)12-m-aperture FLUTE\(\sim\!\)12 could be the right stepping stone for discovering potentially habitable exo-Earth candidates; perhaps to be followed up by a custom mission for a systematic search of biosignatures. This would offer the community a relatively affordable solution to implement a version of the \(\sim\!\)15-m-aperture on-axis Large UV/Optical/IR Surveyor (LUVOIR-A) concept \citep{2019arXiv191206219T}. 
An off-axis FLUTE\(\sim\!\)12 would deliver more exo-Earth candidates than LUVOIR-A even at minimum throughput; and, assuming an optimal throughput, it would yield more than 3 times the \(\sim\!\!\!25\) exo-Earth candidates indicated by the Astro2020 Decadal Survey \citep{2021pdaa.book.....N}, over a 2-year blind-search survey. 

In the shorter term, fluidic shaping could even help implement dedicated missions such as the one suggested by \citet{2025arXiv250913513W}: i.e., a 3-m-aperture space telescope to conduct a relative-astrometry monitoring campaign on \(\alpha\) Centauri \textit{A} and its habitable-zone giant exoplanet candidate directly re-imaged by JWST of late \citep{2025ApJ...989L..22B, 2025ApJ...989L..23S}, in search for (habitable) exomoons.

\newpage

\section{Dual Spectrographs \(\vert\) Habitability to Biosignatures}
\label{ExoSpec}

For a holistic design approach to telescopes, one must also bear in mind that, to conserve the \'etendue (or \(A\Omega\) product) of the optical system, the linear size of traditional bulk-optic instrumentation tends to directly scale with the primary aperture \citep{2006SPIE.6269E..0NB}. 
In this respect, integrated astro-photonics can offer instrumentation with a compact footprint (with on-chip light dispersion), low mass, and that is relatively cheap to fabricate. 
This is why integrated astro-photonics represents the revolutionary pathway to spectrographs, as well as coronagraphs and nulling interferometers, in the long term. 
In particular, the volume of astro-photonic spectrographs is decoupled from the telescope aperture for a given spectral resolving power, operating at the diffraction limit \citep{2019BAAS...51g.270J, Gatkine2019State, 2023JPhP....5d2501J}. 
As such, the rise of astrophotonics' compact yet powerful spectrographs \citep{2019OptPN..30...26N} and coronagraphs \citep{2024SPIE13092E..1TS} can augment the science margins of HWO and ideally fits the scaling potential of fluidic telescopes. 

Along a more conventional pathway, what follows is a preliminary concept of multi-configuration bulk-optic spectrograph tailored for HWO in the medium term. In particular, a refractive dual-configuration integral-field spectrograph that can easily switch between low and moderate spectral-resolving-power (\(R\)) modes in order to characterize both the habitability and the inhabitability of exo-Earths.

\subsection{Baseline Optical Design \& Concept of Operations}

\begin{figure}[ht]
\includegraphics[width=\columnwidth]{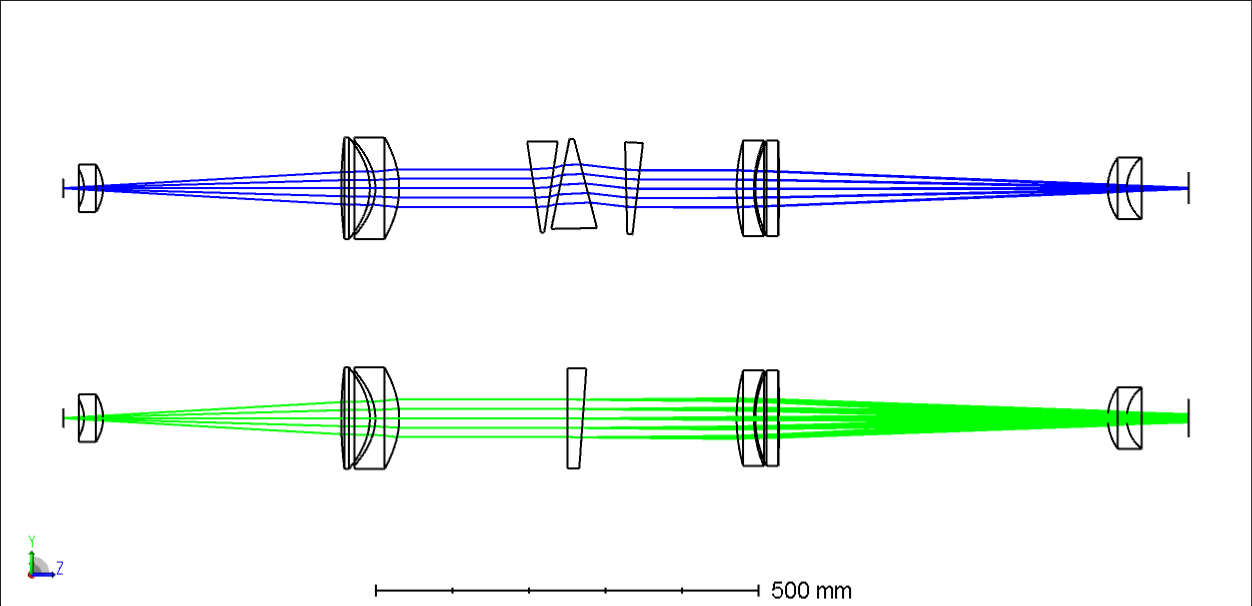}
\caption{\small Dual spectrograph optical design layout (with central-field rays), starting from a design by Qian Gong---in Ansys Zemax OpticStudio. It has fixed front-end (left) and back-end (right) optics, while dispersers are interchangeable. \textit{Top}: low-\(R\!\sim\)140 prismatic mode. \textit{Bottom}: grismatic mode with a moderate \(R\!\sim\)10\(^3\). 
}
\label{Dual Configuration Spectrograph}
\end{figure}

Fig. \ref{Dual Configuration Spectrograph} shows the low-\(R\!\!\!\sim\)140 imaging spectrograph mode (\textit{top}), which is capable of simultaneously taking multiple spectra over an extended field of view, thus producing a spectral data-cube at each integration. The moderate-\(R\) mode (\textit{bottom}) can serve as a single-object (e.g., with a movable slit) or multi-object (e.g., with a micro-shutter array) spectrograph; its ansatz \(R\) was guesstimated at 10\(^3\). 

\clearpage

The hybrid dual-configuration spectrograph in direct-vision geometry of Fig. \ref{Dual Configuration Spectrograph} is designed at a reference wavelength of 760 nm; and it operates over the wavelength range covering at least 600\,--1000 nm. 
The main characteristics of the optical train are the following. 
On the left, it starts with the focus of a square-based lenslet array, illuminated by a centered square field of view, which gets re-imaged to a \(\sim\)40-mm-side square detector.
A field lens and collimating doublet lens group feed the dispersers.  
Finally, there is an imaging doublet followed by a field lens focusing on the detector. 
Dispersers can be mounted on a linear guideway and interchanged via a motorized slider mechanism; collimator and imager can be folded with respect to the dispersers. 

The \(R\!\!\!\sim\)140 mode is suitable for a survey of potentially habitable exo-Earth candidates \citep{2019arXiv191206219T}. 
Its disperser is a double-Amici-style triplet compound prism made of the following materials, found thanks to \citet{2011ApOpt..50.5012H}: \verb|S-LAH60MQ| + \verb|S-NBH58| + \verb|S-LAH60MQ|. 
The \(R\!\!\sim\)10\(^3\) mode aims at looking for clusters of bio-/techno-signatures. 
Its disperser is a grism---i.e., a prism with encoded transmissive diffraction grating---composed of \verb|BSL7Y|, with a groove density of 45 lines/mm. 

The optical performance of the dual spectrograph is basically determined by the detector's pixel pitch, assumed to be 10 \SI{}{\micro\meter}. Its optical aberrations should be such that spectra do not suffer from adjacent spectra cross-talk; which is currently the case, but for the 600\,--1000-nm bandwidth edges.

An alternative bulk-optic solution to this refractive approach---which stems from \citet{2013SPIE.8864E..1OM}---is a reflective multi-configuration spectrograph, with reflective front- and back-end optics for saving throughput---as well as improving luminescence and radiation hardness. In fact, optical transmission can decrease by several percent over the train of refractive lens surfaces, despite the lens anti-reflection coatings. 
The main difference between refractive and reflective solutions lies in the trade-off between the former's design simplicity---and, thus, the implementability, in combination with fold mirrors---against the throughput performance (and practical achromaticity) of the latter.

The dual astrophysical concept of operations leads us to a preliminary trade study for the dual-configuration spectrograph, which is fed by the lenslet array, connected to the module comprising an (extreme) adaptive-optic coronagraphic system---in turn, linked to the optical telescope element. In fact, current coronagraphs are limited to a relative instantaneous bandwidth of \(\sim\)20\(\%\). To start with, let us assume it to be \(\sim\)23\(\%\) with respect to the reference wavelength of \(\sim\)702 nm, to simultaneously cover the three main bands of molecular oxygen (\(O_2\)) shown in the left inset of Fig. \ref{Dual Spectrograph Performance}. 
With this in mind, given various instrumental HWO set-ups, what signal-to-noise ratio (S/N) is achievable as a function of \(R\)? Hereafter is a pilot sample of trade-off analysis on direct exo-Earth spectrography in reflected starlight. 

\newpage

\subsection{Optimal \(R\) Region Trade-Off: Pilot Simulations}

Building upon the noise modeling package by \citet{2019JOSS....4.1387L} and \citet{2016PASP..128b5003R}, let us target a modern exo-Earth at quadrature phase to a Sun-like star 10 parsecs away for 1000 hours via angular differential imaging, equipped with a coronagraphic system and a lenslet-based integral-field spectrograph. 
The main instrumental parameter prescription for the simulations is similar to HWO's \(\sim\)6-m EAC-1 specifications \citep{HabWorlds_SciEn}. 
In this model, signal and noise budgets, as well as the detector's properties, are approximately treated without spatial distribution. 
To maximize the signal-to-noise ratio to detect \(O_2\), we make use of the spectral cross-correlation matched-filter technique with a covariance matrix of the noise \citep{2017ApJ...842...14R}, assuming perfect knowledge of the exo-Earth spectrum as well. 

\begin{figure}[ht]
\centering
\includegraphics[width=\columnwidth]{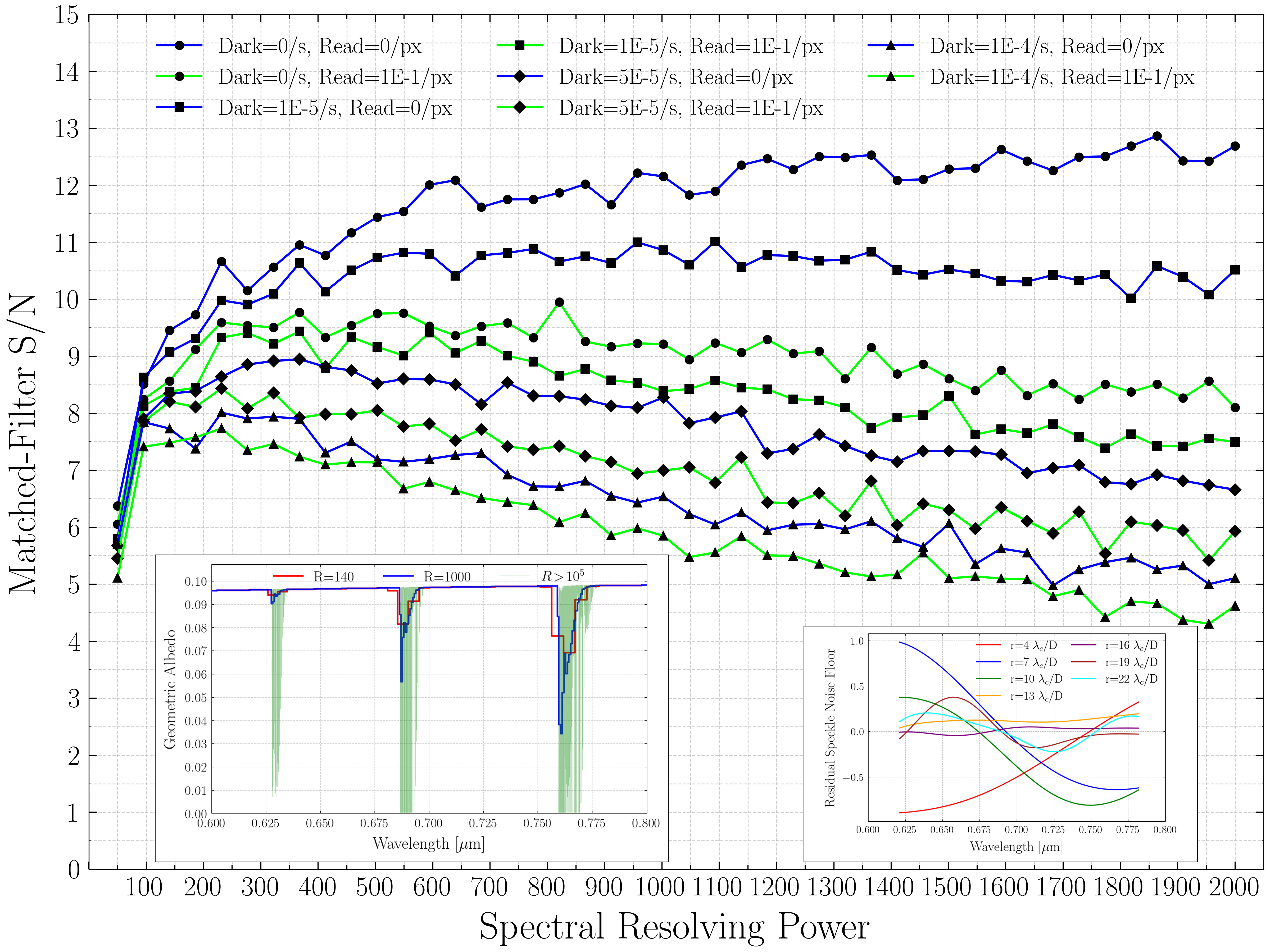}
\vspace{-.786cm}
\caption{\small In the large graph, simulated matched-filter S/N over 1000 hours as a function of \(R\), dark current, and read noise, for a modern exo-Earth orbiting a Sun-like star as observed by EAC-1. The insets show the main ingredients---Earth's (low) \(O_2\) geometric albedo (\textit{left}) and sample residual speckle model (\textit{right})---of the pilot simulations to find the optimal \(R\) region for the spectrograph.
}
\label{Dual Spectrograph Performance}
\end{figure}

Fig. \ref{Dual Spectrograph Performance} (\textit{left inset}) shows \(O_2\)'s geometric albedo spectrum in use \citep{JiWang+2017}, at high \(R\) and degraded at lower, sample values of \(R\) via the Gaussian-convolution re-gridding function of \citet{pyEDITH}. 
Fig. \ref{Dual Spectrograph Performance} (\textit{right inset}) displays samples of the residual quasi-static speckle chromatic noise model extracted via aperture photometry and then rescaled to a target RMS systematic bias of 10\(^{-12}\) from coronagraphic EAC-1 simulations (provided by Roser Juanola-Parramon). 
Fig. \ref{Dual Spectrograph Performance} (\textit{large graph}) shows that dark current and read noise mainly impact observations at higher \(R\) values. Hence, the revolutionary pathway for detectors implies effectively no noise (top curve) and, thus, no penalty for spreading the signal over more pixels for higher-\(R\) spectrographs. 
To this end, superconducting, photon-counting, low-\(R\) energy-resolving microwave kinetic-inductance detectors are promising candidates \citep{2024JATIS..10b5008H}. 

\vspace{-.3cm}

\clearpage

{\bf Acknowledgements.} 
This paper was written as part of NASA's Exoplanet Spectroscopy (alias ExoSpec) Technology Research Collaboration and of FLUTE's team. The latter was mainly supported by a Phase II Study award granted by NASA's Innovative Advanced Concepts (NIAC) Program via the proposal number: \verb|24-24NIACA2-P-0025|. \\
\indent
Enhancements of the codes that produced the graphs and table above were aided by Microsoft Copilot and ChatGPT. \\
For their tips to refine Fig. \ref{Macro-Evolution of Optical Telescopes}, thanks to Jordan D. March\'e II, James M. Lattis, and Wayne H. Osborn (Historical Astronomy Division \(\vert\) American Astronomical Society); to Robert Stack and Peter Ceravolo (Antique Telescope Society).

\bibliography{author.bib}

\end{document}